\documentclass[12pt]{article}
\usepackage{epsfig}

\newcommand{\fig}[1]{Fig.\ref{#1}}

\newcommand{\eqn}[1]{Eq.(\ref{#1})}

\newcommand{\ben}{\begin{enumerate}}
\newcommand{\een}{\end{enumerate}}
\newcommand{\bit}{\begin{itemize}}
\newcommand{\eit}{\end{itemize}}
\newcommand{\bc}{\begin{center}}
\newcommand{\ec}{\end{center}}
\newcommand{\bq}{\begin{equation}}
\newcommand{\eq}{\end{equation}}
\newcommand{\bqa}{\begin{eqnarray}}
\newcommand{\eqa}{\end{eqnarray}}

\newcommand{\nn}{\nonumber}
\newcommand{\nl}{\nonumber\\}

\newcommand{\plaat}[3]{\raisebox{#3pt}{\epsfig{figure=./#1.eps,
width=#2cm}}}

\def\demo{$\Delta\eta\mu \acute{o} \kappa \varrho \iota \tau o \varsigma$}

\begin{document}

\pagestyle{empty}

\begin{flushright}
\end{flushright}

\vspace*{2cm} 

\bc\begin{LARGE} {\bf PHEGAS: a phase-space generator 
for automatic cross-section computation}\\ \end{LARGE}
\vspace*{2cm}

{\large {\bf Costas G.~Papadopoulos } }\\[12pt]
Institute of Nuclear Physics, NCSR \demo, 15310 Athens, Greece
\\[6pt]
E-mail: {\tt Costas.Papadopoulos@cern.ch} 
\\

\vspace*{1cm}

{\bf ABSTRACT}\\[12pt]  \ec

\begin{quote}

A phase-space generation algorithm, capable to
efficiently integrate the squared amplitude of any scattering process, 
is presented.
The algorithm has been implemented in a Monte Carlo program, {\tt PHEGAS},
which, using {\tt HELAC}, a helicity amplitude computational package,
can be used for automatic cross-section computation and event generation.
Results for several scattering processes with four, five and
six particles in the final state are briefly presented.   

\end{quote}

\vspace*{\fill}

July 2000

\vspace*{\fill}

\newpage
\pagestyle{plain}







The study of multi-particle processes, like for instance four-fermion 
production in $e^+e^-$, requires efficient phase-space Monte Carlo generators.
The reason is that the squared amplitude, being a complicated function
of the kinemtaical variables, exhibits strong
variations in specific regions and/or directions of the phase space, 
lowering in a substantial
way the speed and the efficiency of the Monte Carlo integration. A well known
way out of this problem relies on  
algorithms characterized by two main ingredients:
\begin{enumerate}
\item The construction of appropriate mappings of the phase space 
parametrization in such a way that the main variation of the integrand
can be described by a set of almost uncorrelated  variables, and
\item A self-adaptation procedure that reshapes the generated phase-space
density in order to be as much as possible close to the integrand.
\end{enumerate}
Up to now such algorithms have been developed in several cases
to deal with specific processes, like four-fermion~\cite{four-fermion}, four-fermion
plus a photon~\cite{four-fermion+g,minilep2,racoonww} 
and six fermion~\cite{six-fermion} production
in $e^+e^-$ collisions, as well as 
in the framework of general-purpose computational packages like
{\tt CompHEP}~\cite{comphep} and {\tt GRACE}~\cite{grace}.   
It is the aim of this letter
to present a generalized recursive algorithm, together with its implementation,
that can be used for automatic cross-section computation
for any multi-particle process.

In order to construct appropriate mappings we note that  
the integrand, i.e. the squared amplitude,  has a well-defined
representation in terms of Feynman diagrams. It is therefore natural
to associate to each Feynman diagram a phase-space mapping that 
parametrizes the leading variation coming from it. 
To be more specific the contribution 
of tree-order Feynman diagrams to the full amplitude can be factorized
in terms of propagators, vertex factors and external wave functions.
In general, the main source of variation comes from the propagator
factors and therefore our aim is to construct a mapping 
that expresses the phase-space
density in terms of the kinematical invariants that appear
in these propagator factors. 
Since in principle we need
as many mappings as Feynman diagrams for the process under consideration,
we have to appropriately combine them in order to produce
the global phase-space density. A simple and well studied solution
to this problem was suggested some time ago
in reference~\cite{multi}.
Let us represent the normalized phase space-density of a mapping by 
a function 
$g_i(\Phi)$ where $\Phi$ refers to the $(3n-4)$-dimensional phase space,
$n$ being the number of produced particles.
The overall density can be represented by
\[
 g(\Phi)=\sum_{i=1}^{M} \alpha_i g_i(\Phi) 
\]
where 
\[ 0<\alpha_i<1\;\;\;\;\;\sum_{i=1}^{M} \alpha_i=1
\]
and $M$ is the total number of mappings. Since the result of the
integration does not depend on the specific values of $\alpha_i$, 
the so-called {\it a priori} weights,  the latter
can be used to optimize the Monte Carlo integration. A self-adaptation
procedure therefore suggests itself:
during the evaluation of the integral, $\alpha_i$ are repeatedly
redefined~\cite{multi}, so that the variance of the integrand 
is minimized.  
It should be 
mentioned however that other self-adapting approaches can be 
used as well~\cite{Ohl:1999jn}.

In order to describe the construction of the phase-space mappings,
let us consider a typical process in which two incoming particles
produce $n$ outgoing ones. The phase space can be represented by 
\bq
d\Phi_n(P;p_1\ldots,p_n)=(2\pi)^{4-3n}\delta^4\left(\sum_{i=1}^{n} p_i-P\right)
 \prod_{i=1}^n d^4p_i\;
\delta(p_i^2-m_i^2)
\label{phase-space}\eq
where $P=q_1+q_2$ with $q_1$, $q_2$ being the momenta of the
incoming particles.

A well known property of \eqn{phase-space} is that the phase space
can be decomposed as follows
\bqa
 d\Phi_{n} (P;p_1,p_2,\ldots,p_n)&=&
\left(\prod_{i=1}^{m}\frac{dQ_i^2}{2\pi}\right)
d\Phi_{m}(P;Q_1,\ldots,Q_m)
\nl &&
 d\Phi_{n_1}(Q_1;r_1,r_2,\ldots,r_{n_1})
\ldots
 d\Phi_{n_m}(Q_m;s_1,s_2,\ldots,s_{n_m})
\eqa
where the subsets $\{r_1,r_2,\ldots,r_{n_1}\}$ 
up to  $\{s_1,s_2,\ldots,s_{n_m}\}$ 
represent
an arbitrary partition of $\{p_1,p_2,\ldots,p_n\}$.
The above equation can be generalized recursively resulting 
in an arbitrary decomposition of $d\Phi_{n}$.
Feynman graphs can be seen as a realization
of such a decomposition, this latter being identified with a 
sequence of vertices of the graph. 
For instance a three-particle vertex $V=(Q\to Q_1,Q_2)$ in a Feynman diagram
can be seen as part of the phase-space decomposition
\bq 
d\Phi_n = \ldots \frac{d Q_1^2}{2 \pi}\; \frac{d Q_2^2}{2 \pi}\;
d\Phi_2(Q;Q_1,Q_2) \ldots \;\;\;.
\label{example}\eq
The appropriate  sequence of vertices, $\{V_1, V_2, \ldots, V_k\}$ 
can be chosen in such a way that a recursive construction of the phase space
is realized. For instance $V_1$ should contain at least one incoming 
particle whose momentum is known. The rest of the sequence 
is chosen recursively: vertex $V_j$
is characterized by an incoming momentum $Q$ which has already been generated
in one of the $\{V_1,\ldots,V_{j-1}\}$ and outgoing momenta $Q_1$
and $Q_2$ that are generated according to \eqn{example}. 

\begin{figure}[t]
\begin{center}
\mbox{
\epsfig{file=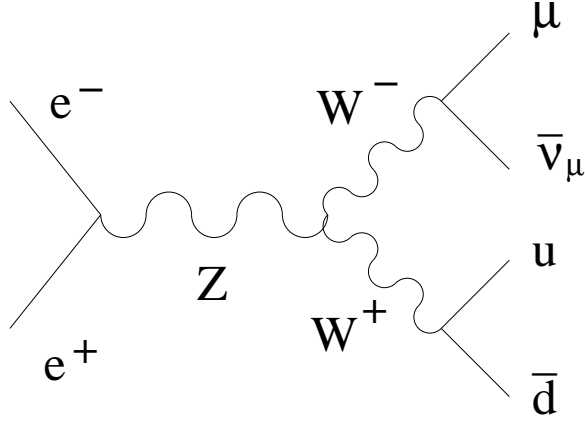,height=6cm,width=8cm}}
\caption[.]{Feynman graph contributing to $e^-\; e^+\to
\mu^-\; \bar{\nu}_\mu\; u\; \bar{d} $.}
\label{graph1}
\end{center}
\end{figure}

As a more illustrative example let us consider the graph
of \fig{graph1} for the process 
\[ 
e^-(q_1)\; e^+(q_2)\to\mu^-(p_1)\; \bar{\nu}_\mu(p_2)\; u(p_3)\; \bar{d}(p_4)
\]
The appropriate sequence of vertices can be chosen as
\[
V_1=(q_1\to -q_2,Q), V_2=(Q\to Q_1,Q_2), 
V_3=(Q_1\to p_1,p_2),V_4=(Q_2\to p_3,p_4)
\]
where $Q, Q_1, Q_2$ are the momenta of $Z, W^-, W^+$ respectively.
In the first vertex, $V_1$, both $q_1$ and $q_2$ are considered known so
this is a mere definition of $Q=q_1+q_2$. 
The rest of the sequence realizes 
the following phase-space decomposition
\bqa
d\Phi_{4} (q_1+q_2;p_1,p_2,p_3,p_4)&=&
d\Phi_{2}(q_1+q_2;Q_1,Q_2)
\nl &&
\frac{dQ_1^2}{2\pi}\frac{dQ_2^2}{2\pi}
\nl &&
d\Phi_{2}(Q_1;p_1,p_2)
\nl &&
d\Phi_{2}(Q_2;p_3,p_4)
\eqa
allowing for the correct treatment of the Breit-Wigner propagators
of $W^\pm$ in terms of the variables $Q_1^2$ and $Q_2^2$.

In the general case one can distinguish two types of 
vertices:
\begin{enumerate}
\item All outgoing momenta involved in the vertex are time-like.
\item One of them is space-like.
\end{enumerate}
It is worthwhile to mention that for $2\to n$ scattering these two cases
are the only possible ones\footnote{Notice that in the present study 
we restrict ourselves to scattering processes whose amplitudes
do not exhibit non-integrable singularities over the available 
phase space.}.  

For the first case the phase space decomposition can be written as 
\bqa
&&\plaat{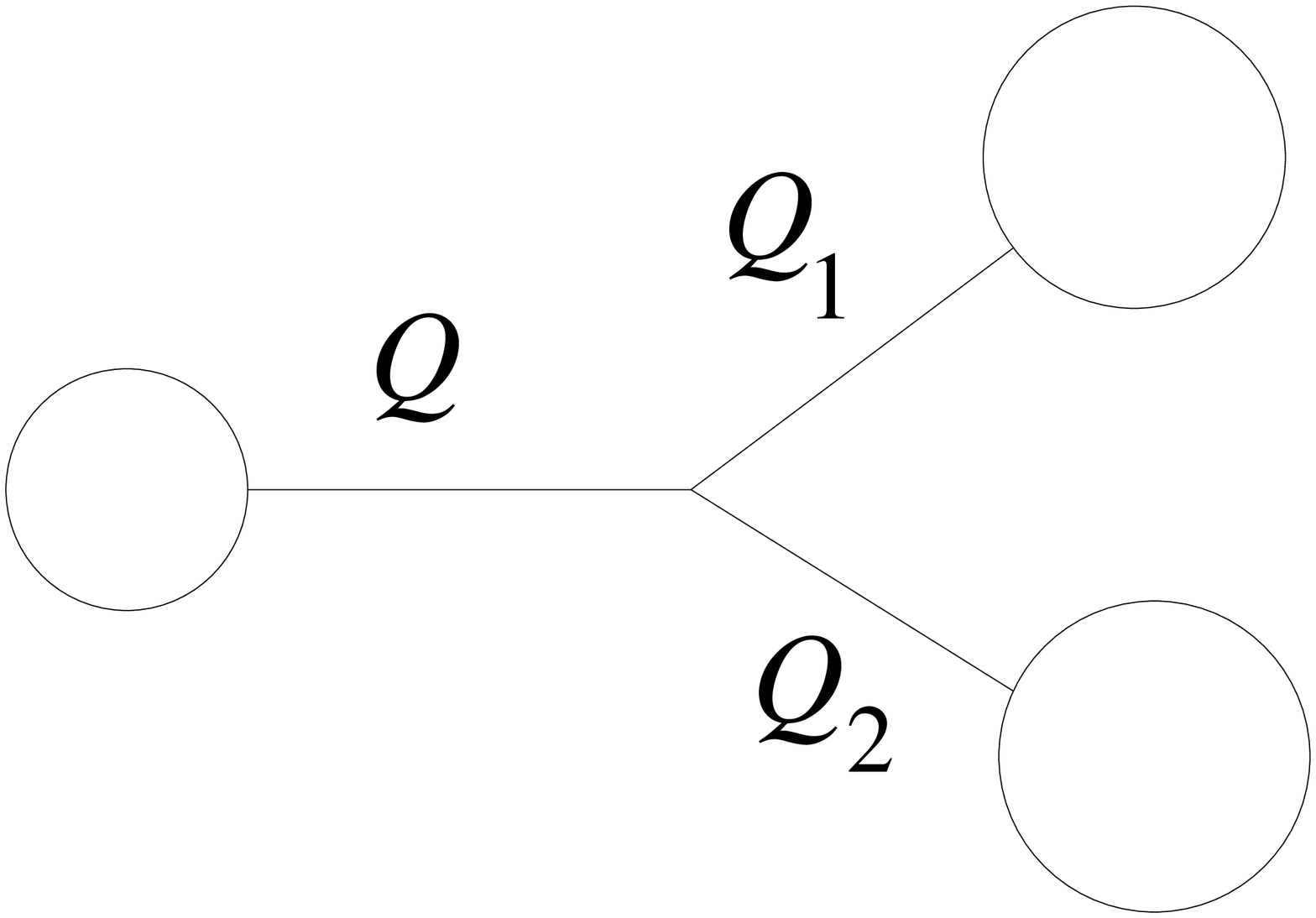}{4}{12}\nl
 d\Phi_n &=& \ldots \frac{d Q_1^2}{2 \pi}\; \frac{d Q_2^2}{2 \pi}\;
d\Phi_2(Q \to Q_1,Q_2) \ldots \nl
&=& \ldots \frac{d Q_1^2}{2 \pi}\; \frac{d Q_2^2}{2 \pi}\;
d \cos\theta \;d \phi\;
 \frac{\lambda^{1/2}(Q^2,Q_1^2,Q_2^2)}{32 \pi^2\; Q^2} \ldots
\nn\eqa
\[ \lambda(x,y,z)=x^2+y^2+z^2-2 xy-2 x z -2 yz \]
with the understanding that whenever $Q_1$ or $Q_2$ represents an external
momentum the corresponding factor $d Q_i^2/2 \pi$ is set to $1$.
Generation can now proceed straightforwardly,
by first generating $Q_1^2$ and $Q_2^2$ using any prescribed
density, as well as $\cos\theta$ and $\phi$
in the rest frame of $Q$. 
Then by using the known momentum $Q$ a boost to the initial frame
can be performed. 
As it is easily seen the first case results to a rather simple
generation algorithm.

The second case is more involved. The phase space is 
decomposed as follows:
\bqa
&&\plaat{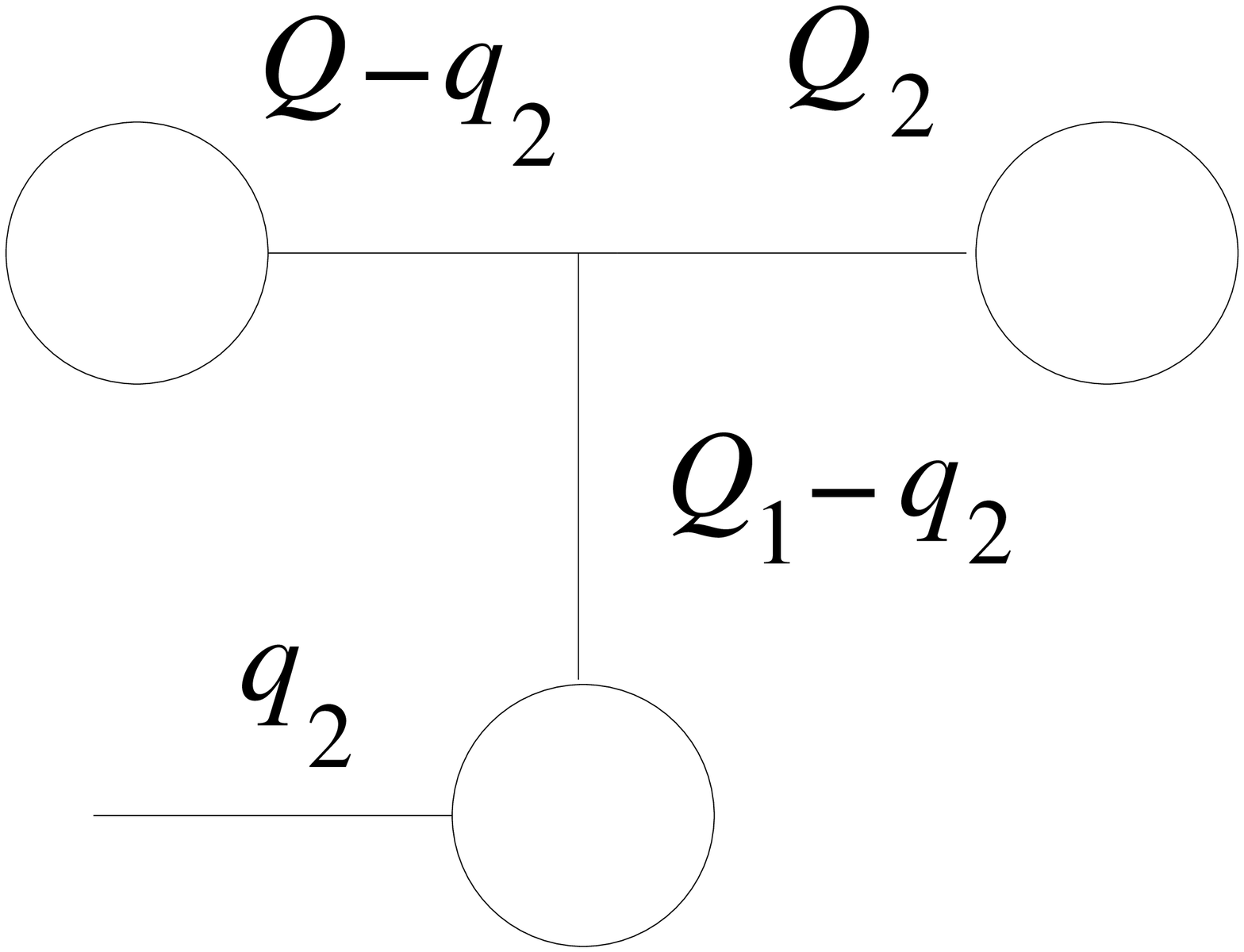}{4}{24}\nl
d\Phi_n &=& \ldots \frac{d Q_1^2}{2 \pi}\; \frac{d Q_2^2}{2 \pi}\;
d\Phi_2(Q \to Q_1,Q_2) \ldots \nl
&=&  \ldots \frac{d Q_1^2}{2 \pi} \;\frac{d Q_2^2}{2 \pi}\;
dt\; d\phi\; \frac{1}{32 \pi^2\; Q\; |\vec{q}_2|}\ldots
\label{t-channel}
\eqa
with
\[ t=(Q_1 - q_2 )^2=
m_2^2+Q_1^2-\frac{E_2}{Q}(Q^2+Q_1^2-Q_2^2)+
\frac{\lambda^{1/2}}{Q}|\vec{q}_2| \cos\theta 
\]
and $(E_2,\vec{q}_2)$ being the incoming momentum $q_2$ in the
rest frame of $Q$.
In order to have an efficient generation according to \eqn{t-channel}
we need to know the limits of the $t-$ and $Q_1^2-$integration:
a detailed presentation of their derivation can be found 
in the Appendix. 

Although in the two cases described so far we used a three-particle vertex
the algorithm can be generalized in a straightforward way in the case
of a four-particle vertex, either
using the three-body phase space explicitly
\[
d\Phi_n = \ldots \frac{d Q_1^2}{2 \pi}\; \frac{d Q_2^2}{2 \pi}\; 
\frac{d Q_3^2}{2 \pi}\;
d\Phi_3(Q \to Q_1,Q_2,Q_3) \ldots 
\]
in the case all momenta are time-like, or using
\[
d\Phi_n = \ldots \frac{d Q_1^2}{2 \pi}\; \frac{d Q_{23}^2}{2 \pi}\; 
d\Phi_2(Q \to Q_1,Q_{23})
\]
followed by 
\[\ldots \frac{d Q_2^2}{2 \pi}\; 
\frac{d Q_3^2}{2 \pi}d\Phi_2(Q_{23} \to Q_2,Q_3) \ldots 
\]
in the case one space-like outgoing momentum is present.

Following the above described algorithm we end up with an expression
for the phase-space density,
\bq
d\Phi_n \to \prod ds_i \; \prod dt_j \;
\prod d\phi_k \prod d\cos\theta_l
\eq
where $s_i$ and $t_j$ refer to the kinematical invariants
entering the propagator factors of the graph and $\phi_k$ and $\cos\theta_l$
represent center-of-mass angles needed to complete the phase space
parametrization. It is now straightforward to generate
$s_i$ and $t_j$ with a probability density given by:
\begin{itemize}
\item
\parbox{5cm}{ $p(x)\sim(x-m^2)^2+m^2 \Gamma^2$}
\parbox[t]{9cm}{for massive unstable particles, like $W^\pm$, $Z$, $t$ .}
\item
\parbox{5cm}{ $p(x)\sim x^{\nu}$}
\parbox[t]{9cm}{for time-like massless propagators,
e.g. $\gamma$, gluons, massless fermions.}
\item
\parbox{5cm}{$p(x)\sim |x|^{\nu}$}
\parbox[t]{9cm}{for space-like massless propagators.}
\end{itemize} 
so that the corresponding propagator factor cancels out in the 
Monte Carlo weight. 
The value of the exponent $\nu$, for $\gamma$ and gluons,
is chosen very close to 1 in order to
account for the leading single-pole behaviour of the squared amplitude
as a result of the gauge cancellations.  

The implementation of this algorithm, called {\tt PHEGAS}, 
is based on and combined with {\tt HELAC}~\cite{helac}
a package that computes any tree-order matrix element.
{\tt HELAC} is based on the Dyson-Schwin\-ger recursive
equations that proved to be superior to the Feynman diagram 
representation for amplitude computation. On the other hand it 
is still an open problem 
how to use Dyson-Schwin\-ger representation to define 
phase-space mappings. 
We have, therefore, implemented an algorithm that
allows the construction of all Feynman diagrams form the {\tt HELAC}
solution of the Dyson-Schwin\-ger equations, in a from suitable
to be used by {\tt PHEGAS}. In fact each Feynman diagram is
represented by a sequence of integer arrays corresponding to 
its vertices. 
The user supplies information concerning the process under investigation
such as the flavours of incoming and outgoing particles as well as
a couple of control parameters as described in reference~\cite{helac},
along with a user-prescribed routine that specifies the desired cuts
on the kinematical variables.  
The {\tt HELAC}-solution of the Dyson-Schwinger equations for the
process under consideration is used to produce the appropriate
representation of all Feynman graphs.
This information is then introduced into {\tt PHEGAS} which produces
phase space points according to the parametrization suggested by
the corresponding mapping as well as the appropriate weight, taking
automatically into account the prescribed phase-space cuts.
The global density is then constructed by computing phase-space
densities for all mappings followed by a multichannel optimization.
The output of the program provides the total cross section as well 
as any kinematical distribution prescribed by the user.

\begin{figure}[t]
\begin{center}
\mbox{
\epsfig{file=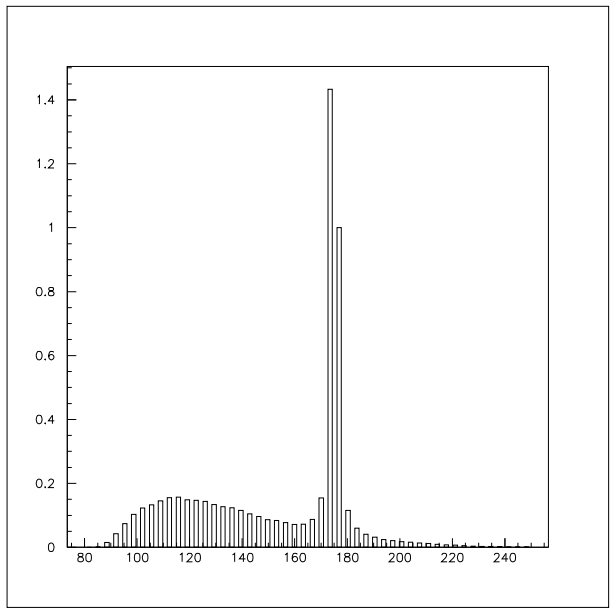,height=8cm,width=8cm}}
\caption[.]{
Differential distribution of 
the invariant masses $m_{b-W^+}$ and $m_{\bar{b}-W^-}$
for the process $gg\to b\bar{b}b\bar{b}W^+W^-$.}
\label{fig2}
\end{center}
\end{figure}

In order to show explicitly the usefulness of the proposed algorithm
we consider the following  typical examples of cross-section computation.
\begin{itemize}
\item  $e^- e^+ \to \mu \, \bar{\nu}_\mu\, u \,\bar{d}$ \\
This is a well studied process within four-fermion physics at LEP2.
We present here results form {\tt PHEGAS/HELAC} in comparison with 
results from {\tt EXCALIBUR}~\cite{exca,four-fermion}. 
They are summarized in the following table:

\bc
\begin{tabular}{|l|c|c|c|c|}
\hline
& MC points & result  & error  & efficiency 
\\
& $w>0$ & (fb) & (fb) & (\%)
\\
\hline
{\tt PHEGAS/HELAC} &  1510700  & 608.64 & 0.61 & 3.5
\\
\hline
{\tt EXCALIBUR} & 1574175 & 608.22 & 0.57 & 3.6
\\
\hline
\end{tabular}
\ec
where we have used $2\times 10^6$ MC points, at $\sqrt{s}=190$ GeV, and
a fixed width prescription for internal unstable-particle propagators
in both programs and identical input parameters. In the last column of the table
the efficiency of the generator is given. The efficiency of an event-generator 
is defined as the ratio of the mean to the maximum Monte Carlo weight and it is
also related to the number of the unweighted events: for instance
in the above run a sample of $2\times 10^6\times0.035\sim70000$ unweighted
events would have been produced.

Moreover the following set of cuts has been applied:
\[
M_{u,\bar{d}}>10\mbox{GeV},\;\;
|\cos\theta(u(\bar{d}),
\mbox{beam})|<0.9,\;\;
\cos\theta(u,\bar{d})<0.9,\;\;
E_{u(\bar{d})} > 20\mbox{GeV} .
\]
Both programs are equally fast and the run of $2\times 10^6$ MC points
costs no more than a few {\tt CPU} minutes on {\tt DXPLUS@cern.ch}.
	
\item  $e^- e^+ \to \mu\, \bar{\nu}_\mu \,u \,\bar{d}\,\gamma $ \\
In order to demonstrate the ability of {\tt PHEGAS/HELAC} to deal with
more complicated processes we give here results on four-fermion plus a 
gamma
production. The results compare very well with the results presented
in reference~\cite{minilep2} form {\tt WRAP}~\cite{minilep2} and 
{\tt RACOONWW}~\cite{racoonww}
as is is shown in the following table:
\bc
\begin{tabular}{|c|c|c|c|}
\hline
 $e^-e^+\to$    & {\tt WRAP}  & {\tt RACOONWW}  & {\tt PHEGAS/HELAC}
\\
\hline
 $u\bar{d}\mu^-\bar{\nu}_\mu\gamma $  & 75.732(22) & 75.647(44) & 75.683(66)
\\
\hline
 $u\bar{d}e^-\bar{\nu}_e\gamma $  & 78.249(43) & 78.224(47) & 78.186(76)
\\
\hline
 $\nu_\mu\mu^+ \tau^-\bar{\nu}_{\tau}\gamma $  & 28.263(9) & 28.266(17) & 28.296(22)
\\
\hline 
 $\nu_\mu\mu^+ e^-\bar{\nu}_e\gamma $  & 29.304(19) & 29.276(17) & 29.309(25)
\\
\hline
 $u\bar{d} s\bar{c}\gamma $  & 199.63(10) & 199.60(11) & 199.75(16)
\\
\hline
\end{tabular}
\ec
We refer to reference~\cite{minilep2} for details on parameters and cuts used
for this computation, as well as an extensive comparison among the
three generators based on differential distributions, which shows a very good
technical agreement. 

\item  $g\, g \to b\, \bar{b}\, b\, \bar{b}\, W^- W^+ $  \\
The reason we have chosen such a process is twofold: in first place
this is a challenging process, from a computational point of view, 
and secondly this is a nice example to demonstrate the ability
of {\tt PHEGAS/HELAC} to deal with QCD processes. Moreover its study is important
as a background of $t\bar{t} H$ production~\cite{lhc}.
The results of the computation are summarized as follows:
\bc
\begin{tabular}{|c|c|c|c|c|}
\hline
 MC points & result  & error  & efficiency & efficiency
\\
 $w>0$ & (fb) & (fb) & (\%) & $w>0$  (\%)
\\
\hline
 99442  & 4.716 & 0.024 & 3.3 & 33
\\
\hline
\end{tabular}
\ec
The results refer to an energy $\sqrt{s}=500$ GeV and to $1\times 10^6$ MC points.
To give an idea of the complexity of the computation,  
the number of Feynman graphs for this process is {\bf 960},
without taking into account electroweak contributions
from $Z$ and $\gamma$ intermediate states.
Parameters used are $g_{QCD}=1$, $m_{top}=175$ GeV and $\Gamma_{top}=1.5$ GeV.
Moreover the following set of cuts has been applied:
\[
M_{q,q^\prime}>20\mbox{GeV},\;\;
E_{q}>20\mbox{GeV},\;\;
|\cos\theta(q,\mbox{beam})|<0.9,
\]
where $q,q^\prime$ refer to any quark or anti-quark of the final state.
Finally in \fig{fig2} we show the distribution of the invariant masses
of $b-W^+$ and $\bar{b}-W^-$ pairs, exhibiting the expected peak at 
$m_{top}$ along with the non-resonant QCD corrections.
\end{itemize}

In conclusion {\tt PHEGAS/HELAC} provides an automatic and efficient 
computational
framework to perform cross section evaluation and event generation for 
arbitrary scattering processes.


\vspace*{2cm}

{\large\bf Appendix}

\vspace*{12pt}

In this appendix we describe the limits on $t$ and $Q_1^2$.
The expression for the $t$ invariant is given by 
\[ t=(Q_1 - q_2 )^2=
m_2^2+Q_1^2-\frac{E_2}{Q}(Q^2+Q_1^2-Q_2^2)+
\frac{\lambda^{1/2}}{Q}|\vec{q}_2| \cos\theta 
\]
with 
\[ \lambda\equiv\lambda(Q^2,Q_1^2,Q_2^2)=Q^4+Q_1^4+Q_2^4-2 Q_1^2 Q^2-2
Q_2^2Q^2-2Q_1^2Q_2^2
\]

The limits for $t$ can be found by maximizing~(minimizing) 
$t_\pm$ given by
\[
t_\pm=m_2^2+Q_1^2-\frac{E_2}{Q}(Q^2+Q_1^2-Q_2^2)\pm 
\frac{\lambda^{1/2}}{Q}|\vec{q}_2|
\]
In order to find the maximum of $t_+$ we study the function 
$\partial t_{+}/\partial Q_1^2$ in the region
$Q^2_{1,min}<Q^2_1<(Q-Q_2)^2$. Since 
\[ 
\frac{\partial^2 t_{+}}{\partial (Q_1^2)^2}
 = - 4 Q^2Q_2^2\lambda^{-3/2} \frac{|\vec{q}_2|}{Q} \le 0
\] 
and 
\[ 
\partial t_{+}/\partial Q_1^2 |_{Q_1^2=(Q-Q_2)^2} \to -\,\infty 
\]
we just consider two cases ($|\vec{q}_2|\ne 0$):
\begin{enumerate}
\item $\partial t_{+}/\partial Q_1^2 |_{Q_1^2=Q^2_{1,min}} <0$ in which case
$t_{max}=t_{+,max}=t_{+}(Q_1^2=Q^2_{1,min})$, and
\item $\partial t_{+}/\partial Q_1^2 |_{Q_1^2=Q^2_{1,min}} > 0$ in which case
one can easily derive $t_{max}=t_{+}(Q_1^2=x_{-})$ with
\[
x_-=Q^2+Q_2^2-2\;Q\; Q_2 \frac{1-E_2/Q}{\sqrt{\alpha}}\;,\;\; \alpha
=\left(1-\frac{E_2}{Q}\right)^2-\left(\frac{|\vec{q}_2|}{Q}\right)^2 >0
\]
\end{enumerate}
Following the same reasoning we find for the lower limit on $t$
that
\begin{enumerate}
\item $\partial t_{-}/\partial Q_1^2 |_{Q_1^2=Q^2_{1,min}} > 0$ in which case
$t_{min}=t_{-,min}=t_{-}(Q_1^2=Q^2_{1,min})$, and
\item $\partial t_{-}/\partial Q_1^2 |_{Q_1^2=Q^2_{1,min}} < 0$ in which case
one can easily derive $t_{min}=t_{-}(Q_1^2=x_{+})$ with
\[
x_+=Q^2+Q_2^2+2\;Q\; Q_2 \frac{1-E_2/Q}{\sqrt{\alpha}}
\]
\end{enumerate} 
The limits for the $Q_1^2$-integration for given $t$ can now be fixed by the
condition $|\cos\theta|\le 1$ or equivalently
\[\Pi(Q_1^2)\le 0 \]
with
\[
\Pi(Q_1^2)=\left(t -Q_1^2 -m_2^2
+\frac{E_2}{Q}(Q^2+Q_1^2-Q_2^2) \right)^2 -
\left(\frac{|\vec{q}_2|}{Q}\right)^2\lambda
\]
If $y_1\le y_2$ are the two roots of the polynomial $\Pi(Q_1^2)$
then we have
\ben
\item For $ a > 0 $ , $ y_- < Q_1^2 < y_+ $, with $y_-=\max(y_1,Q^2_{1,min})$
and $y_+=\min(y_2,Q^2_{1,max})$
\item For $ a < 0 $ we have 
to satisfy two conditions
$Q_1 ^2 < y_1 \;\;\mbox{or}\;\; y_2 < Q_1 ^2 $ 
and $ Q_{1,min}^2 < Q_1^2 < Q_{1,max}^2 $
\een

\end{document}